\documentclass[useAMS,usenatbib]{mn2e}
\usepackage{aas_macros,graphicx,times,multirow}

\usepackage{graphicx}
\usepackage{standalone}
\usepackage{subfig}
\usepackage{multirow}
\usepackage{array}

\newcommand{\xmm}{{\em XMM-Newton}}
\newcommand{\chan}{{\em Chandra}}

\title[VLT observations of $\gamma$-ray pulsars]{Searching for the optical counterparts of two young $\gamma$-ray pulsars\thanks{Based on observations collected at the European Organisation for Astronomical Research in the Southern Hemisphere under ESO programme 095.D-0328(B)}}
\author[R. P. Mignani, et al. ]
{\parbox{\textwidth}{R. P. Mignani$^{1,2}$\thanks{E-mail: mignani@iasf-milano.inaf.it}, 
V. Testa$^{3}$,
M. Marelli$^{1}$,
A. De Luca$^{1,4}$,
M. Pierbattista$^{5}$,
M. Razzano$^{6}$,
D. Salvetti$^{1}$,
A. Belfiore$^{1}$,
A. Shearer $^{7}$,
P. Moran $^{7}$
} 
\\ \\
$^{1}$ INAF - Istituto di Astrofisica Spaziale e Fisica Cosmica Milano, via E. Bassini 15, 20133, Milano, Italy\\
$^{2}$ Janusz Gil Institute of Astronomy, University of Zielona G\'ora, Lubuska 2, 65-265, Zielona G\'ora, Poland \\
$^{3}$ INAF - Osservatorio Astronomico di Roma, via Frascati 33, 00040, Monteporzio, Italy \\
$^{4}$ INFN - Istituto Nazionale di Fisica Nucleare, sezione di Pavia, via A. Bassi 6, 27100, Pavia, Italy \\
$^{5}$ Maria Curie-Sklodowska University, Department of Astrophysics and Theory of Gravity, ulica Radziszewskiego 10, 20-031 Lublin, Poland\\
$^{6}$ Istituto Nazionale di Fisica Nucleare, Sezione di Pisa, I-56127 Pisa, Italy \\
$^{7}$ Centre for Astronomy, National University of Ireland, Newcastle Road, Galway, Ireland
}

\begin{document}

\date{Accepted 1988 December 15. Received 1988 December 14; in original form 1988 October 11}

\pagerange{\pageref{firstpage}--\pageref{lastpage}} \pubyear{2002}

\maketitle

\label{firstpage}

\begin{abstract}
We report on the first deep optical observations of two $\gamma$-ray pulsars, both among the very first discovered by the {\em Fermi} Gamma-ray Space Telescope.  The two pulsars are the radio-loud PSR\,  J1907+0602 in the TeV pulsar wind nebula (PWN) MGRO\, J1908+06 and the radio-quiet PSR\, J1809$-$2332 in the "Taz" radio/X-ray PWN. These pulsars are relatively young
and energetic
and have been both detected in the X-rays by \xmm, which makes them viable targets for optical observations. We observed the pulsar fields in the B and V bands with the Very Large Telescope (VLT) in June/July 2015 to search for their optical counterparts. Neither of the two pulsars has been detected down to $3\sigma$ limiting magnitudes of $m_{\rm v} \sim 26.9$ and $m_{\rm v} \sim 27.6$ for PSR\,  J1907+0602 and PSR\, J1809$-$2332, respectively. We discuss these results in the framework of the multi-wavelength emission properties of pulsars.
\end{abstract}


\section{Introduction}

Pulsars are rapidly rotating isolated neutron stars, powered by their rotational energy (see Kaspi \& Kramer 2016 for a recent review). Prevalently observed at radio wavelengths, they are also detected at X and $\gamma$-ray energies and in the optical band, where they are challenging targets owing to their intrinsic faintness (Mignani 2011). While in the 1980s/1990s pulsar searches in the optical were mainly driven by their X-ray detection, since the launch of the {\em Fermi} Gamma-ray Space Telescope in 2008 the wealth of pulsar $\gamma$-ray detections (see, Caraveo 2014 and Grenier \& Harding 2015 for recent reviews) have spurred their search both at X-ray and optical wavelengths. With over 200 $\gamma$-ray pulsars now identified by {\em Fermi}\footnote{See {\texttt https://confluence.slac.stanford.edu/display/GLAMCOG/} for a continually updated list.}, 
the number of those detected in the optical (or with at least  a candidate optical counterpart) is still tiny (Abdo et al.\ 2013; Moran et al.\ 2013) owing to the paucity of sensitive optical observations (see, e.g. Mignani et al.\ 2016a for a summary). Recently, we detected a candidate optical counterpart to the middle-aged $\gamma$-ray pulsar PSR\, J1741$-$2054 (Mignani et al.\ 2016b), with the ESO Very Large Telescope (VLT).  In the same run, we observed other two $\gamma$-ray pulsars discovered by {\em Fermi}, PSR\,  J1907+0602  and  PSR\, J1809$-$2332, as part of a dedicated pilot survey. The characteristics of these two pulsars are summarised in Table \ref{psr}.

\begin{table}
\begin{center}
\caption{Coordinates, proper motion, position reference epoch (Kerr et al.\ 2015) for the two {\em Fermi} pulsars discussed in this work, together with their spin period P$_{\rm s}$, period derivative $\dot{P}_{\rm s}$, and inferred values of the characteristic age $\tau_c \equiv P_{\rm s}/ 2 \dot{P}_{\rm s}$, rotational energy loss $\dot{E}_{\rm rot}$ and surface dipolar magnetic field $B_{\rm s}$. The latter two values have been derived from the standard  formulae $\dot{E}_{\rm rot} = 4 \times 10^{46}  \dot{P}_{\rm s}/P_{\rm s}^{3}$ erg s$^{-1}$ and $B_{\rm s} = 3.2 \times 10^{19} \sqrt{P_{\rm s} \dot{P}_{\rm s}}$ G, derived by assuming for the neutron star a moment of inertia $I = 10^{45}$ g cm$^{2}$ (e.g., Kaspi \& Kramer 2016).  The values have been obtained from the ATNF pulsar catalogue (Manchester et al.\ 2005).}
\label{psr}
\begin{tabular}{lll} \hline
		   & PSR\,  J1907+0602 & PSR\,  J1809$-$2332     \\  \hline
 $\alpha$ (J2000) & $19^{\rm h}  07^{\rm m} 54\fs76$ (0\fs05) & $18^{\rm h}  09^{\rm m} 50\fs249$ (0\fs030) \\
 $\delta$ (J2000) & $ +06^\circ 02\arcmin 14\farcs6$ (0\farcs7) & $-23^\circ 32\arcmin 22\farcs67$ (0\farcs10) \\
$\mu_{\alpha}$ (mas yr$^{-1}$) &  - & $+12\pm8$  \\
$\mu_{\delta}$  (mas yr$^{-1}$) & - & $-24\pm6$   \\
 Epoch  (MJD) & 55555 & 55555 \\
 P$_{\rm s}$ (s) & 0.106 & 0.146 \\
 $\dot{P}_{\rm s}$ ($10^{-14}$ s s$^{-1}$) & 8.68 & 3.44 \\
 $\tau_c$ (kyr)   &  19.5 & 67.6 \\
 $\dot{E}_{\rm rot}$ ($10^{36}$ erg s$^{-1}$) & 2.8& 0.43\\
 $B_{\rm s}$ ($10^{12}$ G) & 3.08 & 2.27 \\ \hline
\end{tabular}
\end{center}
\end{table}

PSR\,  J1907+0602 was discovered as a $\gamma$-ray pulsar during a blind search for pulsations in unidentified {\em Fermi}-LAT sources (Abdo et al.\ 2009a; 2010). 
These characteristics make PSR\,  J1907+0602 quite similar to the slightly younger (11.2 kyr) Vela pulsar (Manchester et al.\ 2005), one of the historical $\gamma$-ray pulsars (e.g., Abdo et al.\ 2009b). Very faint radio pulsations from PSR\,  J1907+0602  were  detected with the Arecibo telescope at 1.5 GHz (Abdo et al.\ 2010; Ray et al.\ 2011), soon after its discovery as a $\gamma$-ray pulsar.  The radio dispersion measure (DM=82.1$\pm$1.1 cm$^{-3}$ pc) puts the pulsar at a nominal distance of 3.2$\pm$0.6 kpc (Abdo et al.\ 2010), according to the model of the Galactic free electron density  (Cordes \& Lazio 2002).  This would make PSR\,  J1907+0602 one of the faintest known radio pulsars, with a 1.4 GHz radio luminosity of 0.035 mJy kpc$^{2}$.   No radio parallax measurement has been obtained for this pulsar. PSR\,  J1907+0602  has been searched for but not detected in a pulsar survey at 34 MHz (Maan \& Aswathappa 2014).  Both the DM distance and spin-down age suggest that PSR\,  J1907+0602 was probably born at the centre of the supernova remnant (SNR) G40.5$-$0.5 (Abdo et al.\ 2010). After a preliminary detection by \chan\  (Abdo et al.\ 2010; Marelli et al.\ 2011), the pulsar has been observed by \xmm\ (Abdo et al.\ 2013). No X-ray pulsations have been detected yet. PSR\,  J1907+0602 is likely associated with a pulsar wind nebula (PWN) detected at TeV energies by MILAGRO, HESS, VERITAS, and HAWC (Abdo et al.\ 2010; Abeysekara et al.\ 2016). 

PSR\, J1809$-$2332 is older than PSR\,  J1907+0602,
ideally half way between the young, Vela-like pulsars and the middle-aged ones (100 kyr--1 Myr).  
Like PSR\,  J1907+0602,  PSR\, J1809$-$2332 has been identified as a $\gamma$-ray pulsar by the {\em Fermi}-LAT during a blind search for pulsations  from unidentified sources (Abdo et al.\ 2009a). The LAT source (3FGL\, J1809.8$-$2332; Acero et al.\ 2015) associated with the pulsar is identified with the {\em Compton}/EGRET source 3EG\, J1809$-$2338 (Hartmann et al.\ 1999). The latter was found to be spatially coincident with the dark nebula Lynds 227 and a PWN candidate detected in the X rays by {\em ASCA} (Oka et al.\ 1999).   The PWN, later dubbed the "Tasmanian devil" ("Taz" for short),  was then observed by \chan\  (Braje et al.\ 2002; Roberts \& Brogan 2008), which also resolved the point-like source CXOU\, J180950.2$-$233223 that Abdo et al.\ (2009) identified with the PSR\, J1809$-$2332 X-ray counterpart. The "Taz" PWN is also detected  in radio and located within the shell SNR G7.5$-$1.7 (Roberts \& Brogan 2008).  PSR\, J1809$-$2332  has been also observed by \xmm\ (Marelli et al.\ 2011) but X-ray pulsations have not been detected yet. The pulsar was searched for radio emission (Ray et al.\ 2011) but it was not detected down to a flux limit of 26 $\mu$Jy.  A pulsar search at 34 MHz (Maan \& Aswathappa 2014) also resulted in a non detection. Since PSR\, J1809$-$2332 is not detected in radio there is no direct measurement of its distance. This is estimated to be 1.7$\pm$1.0 kpc from the distance to the dark nebula Lynds 227, which Oka et al.\ (1999) associated with the "Taz" PWN.  Using \chan, Van Etten et al.\ (2012) measured a proper motion 
for PSR\, J1809$-$2332, which confirms the association with the SNR G7.5$-$1.7, as proposed by Roberts \& Brogan (2008).

The optical emission of young pulsars  ($\tau_{\rm c} \la 0.1$ Myr) is ascribed to synchrotron emission  from energetic  electrons in  the  pulsar magnetosphere (e.g., Pacini \& Salvati 1983) and the spectrum is characterised by a power-law (PL). For older pulsars, the thermal emission from the cooling neutron star surface also contributes to the optical emission and the spectrum is the combination of both a PL and a Rayleigh-Jeans (see, e.g. Mignani 2011 for a review). 
Nothing is known about the optical emission properties of PSR\,  J1907+0602 and  PSR\, J1809$-$2332.
Recently, Brownsberger \& Romani (2014) carried out observations of the two pulsar fields in H$_{\alpha}$ with the 4.2m SOAR telescope to search for bow-shock nebulae.  No wide-band imaging observations of the pulsar fields  have been performed, though.  

Here, we report the results of our VLT observations of PSR\,  J1907+0602 and  PSR\, J1809$-$2332, the first carried out with a 10m-class telescope. In Sectn.\ 2 we describe the observations and data analysis, whereas we present and discuss the results in Sectn.\ 3 and 4, respectively.

\begin{figure*}
\centering
\begin{tabular}{cc}
\subfloat[PSR\, J1809-2332]{\includegraphics[width=8cm,bb=0 0 796 791,clip=]{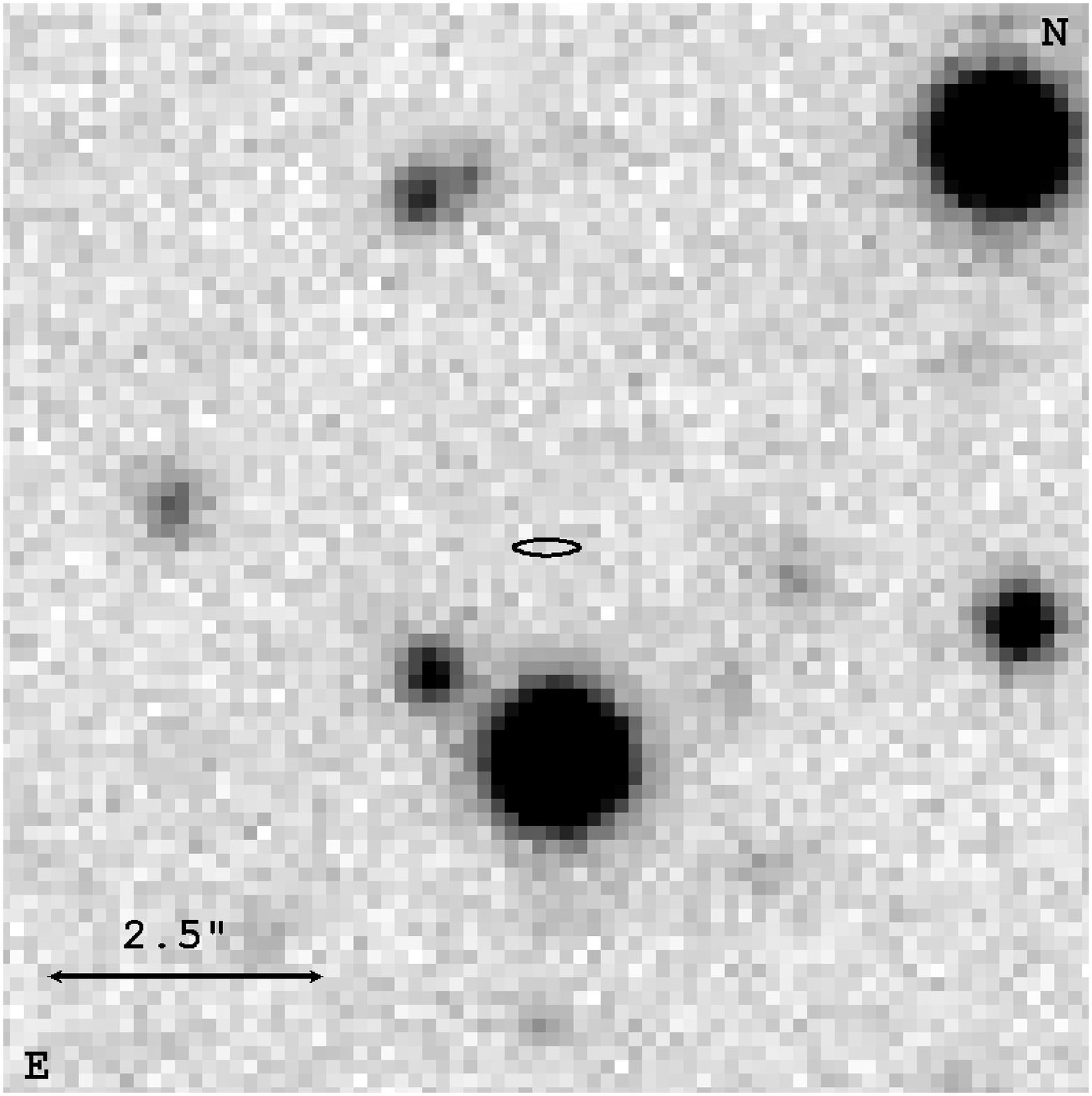}} 
\subfloat[PSR\, J1907+0602]{\includegraphics[width=8cm,bb=0 0 796 791,clip=]{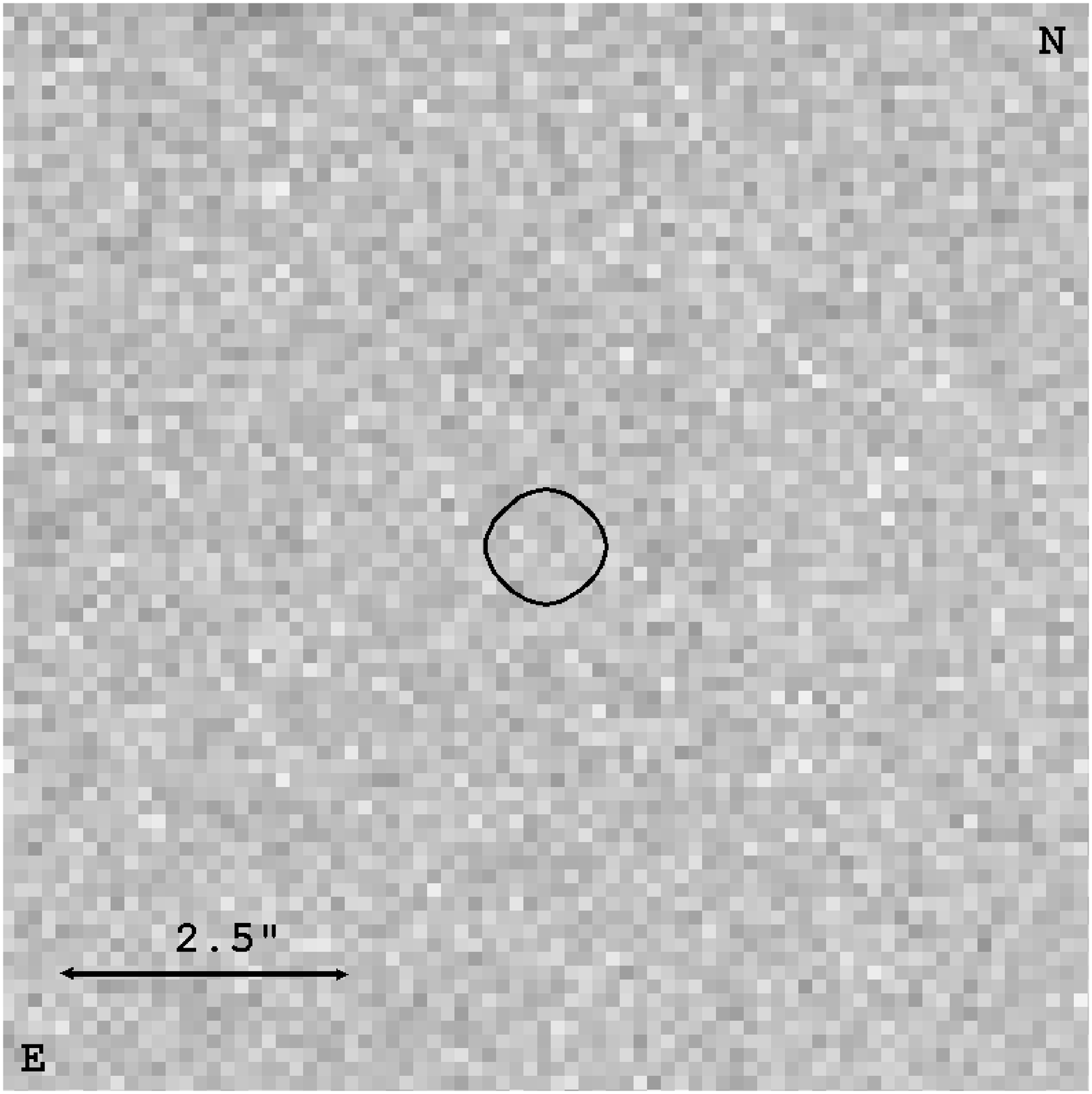}} 
\end{tabular}
\caption{\label{fc} 
$10\arcsec \times 10\arcsec$ VLT/FORS2 $v_{\rm HIGH}$-band images of the pulsar fields. 
The pulsar positions determined by \chan\ (Kerr et al.\ 2015) are marked by the ellipses. The size of the ellipses accounts for statistical uncertainties only and not for the systematic uncertainty associated with the astrometry calibration of the VLT images ($\sim$0\farcs1). The PSR\, J1809$-$2332 position has been corrected for the pulsar proper motion (Van Etten et al.\ 2012). Quite surprisingly, in the case of  PSR\,  J1907+0602 no stars are detected within the entire  $10\arcsec \times 10\arcsec$  region around the reference pulsar position.
}
\end{figure*}

\section{Observations and data reduction}

The two pulsars were observed in service mode in June 20 and July 17, 19 2015 (PSR\, J1809$-$2332) and June 16, 20 2015 (PSR\,  J1907+0602) with the VLT and the second FOcal Reducer and low dispersion Spectrograph  (FORS2;
Appenzeller  et  al.\ 1998).
The camera was used in imaging mode and equipped with its default MIT detector, a mosaic of two 4k$\times$2k CCD chips aligned along the long axis, optimised  for wavelengths  longer  than 6000  \AA.   With the FORS2 high-resolution collimator,   the  detector  has  a  pixel   scale  of  0\farcs125 (2$\times$2 binning)
and a projected field--of--view (FOV) of 4$\farcm25  \times 4\farcm25$.
However,  owing to vignetting produced by the camera optics with the high-resolution collimator, the effective sky coverage is smaller than the nominal FOV, and is larger for the upper CCD chip (4$\farcm25  \times 2\farcm12$).   
The observations were executed with the standard low-gain and fast read-out mode
and through the high-throughput  $b_{\rm HIGH}$ ($\lambda=4400$ \AA;  $\Delta \lambda=1035$\AA)  and $v_{\rm HIGH}$ ($\lambda=5570$ \AA;  $\Delta \lambda=1235$\AA) filters.  The target positions were placed at the nominal aim point of FORS2, close to the lower edge of the upper CCD chip. To allow for  cosmic-ray removal and reduce the impact of bright star saturation, we obtained  sequences of short exposures (180 s each). The total integration time is different for the two pulsar fields.  For PSR\,  J1907+0602, we collected 5400 s and  4500 s in the  $b_{\rm HIGH}$ and $v_{\rm HIGH}$ filters, respectively.   For PSR\, J1809$-$2332, just a few sequences out of the planned ones were executed, amounting to a total integration time of 1080 s ($b_{\rm HIGH}$) and 2700 s ($v_{\rm HIGH}$) only.
All exposures  were taken  in  dark time  and under  clear sky conditions, with average airmass $\sim 1.19$ ($v_{\rm HIGH}$)  and $\sim 1.3$ ($b_{\rm HIGH}$) for PSR\,  J1907+0602 and $\sim 1.04$ ($b_{\rm HIGH}$) and $\sim$1.1 ($v_{\rm HIGH}$) for PSR\, J1809$-$2332.   The average image quality, as measured directly on the frames, was  $\sim$0\farcs9 and 0\farcs5 in the $b_{\rm HIGH}$ and $v_{\rm HIGH}$ bands, respectively for PSR\, J1809$-$2332, whereas for PSR\,  J1907+0602 it was $\sim$0\farcs8 in both filters.

We reduced the data (bias  subtraction and  flat--fielding) using tools in the {\sc IRAF}\footnote{IRAF is distributed by the National Optical Astronomy Observatories, which are operated by the Association of Universities for Research in Astronomy, Inc., under cooperative agreement with the National Science Foundation.} package {\sc ccdred}. Per each band, we aligned and average-stacked the reduced  science images with the  {\tt  drizzle} task in {\sc IRAF}, applying a $\sigma$ clipping to filter  out hot/cold pixels and cosmic ray hits.  
We applied the  photometric calibration by using  the FORS2 night  zero points and the computed atmospheric extinction coefficients for the Paranal Observatory\footnote{\texttt{www.eso.org/observing/dfo/quality/FORS2/qc/qc1.html}}.  
We computed the astrometry calibration using the {\em wcstools}\footnote{\texttt{http://tdc-www.harvard.edu/wcstools}} suite of programs and reference stars from the GSC2.3 (Lasker et al.\ 2008).  We obtained mean residuals of $\la 0\farcs1$ in the radial direction, using up to 30 non-saturated GSC2.3 stars, evenly distributed in the field of view but selected to avoid the vignetted regions of the detector.  To this value we added in quadrature the uncertainty of the image registration  on the GSC-2.3 reference frame  ($\sim$ 0\farcs11) and  the 0\farcs15  uncertainty on the link of the GSC2.3 to the International Celestial Reference Frame. We ended up with an overall accuracy of $\sim$0\farcs2 on our absolute astrometry. Thanks to the pixel scale of the FORS2 images (0\farcs125), the uncertainty on the centroids of the reference stars is negligible.

\begin{figure*}
\centering
\begin{tabular}{cc}
\subfloat[PSR\, J1809-2332]{\includegraphics[width=16cm, bb=0 5 1024 518, clip=]{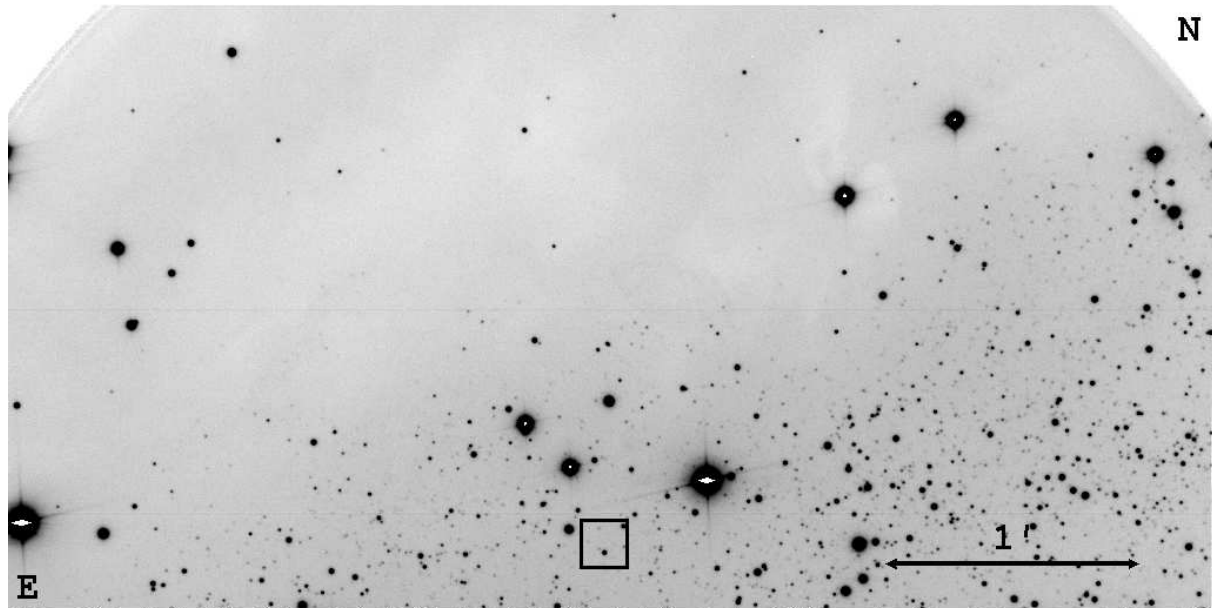}} \\
\subfloat[PSR\, J1907+0602]{\includegraphics[width=16cm, bb=0 5 1024 518, clip=]{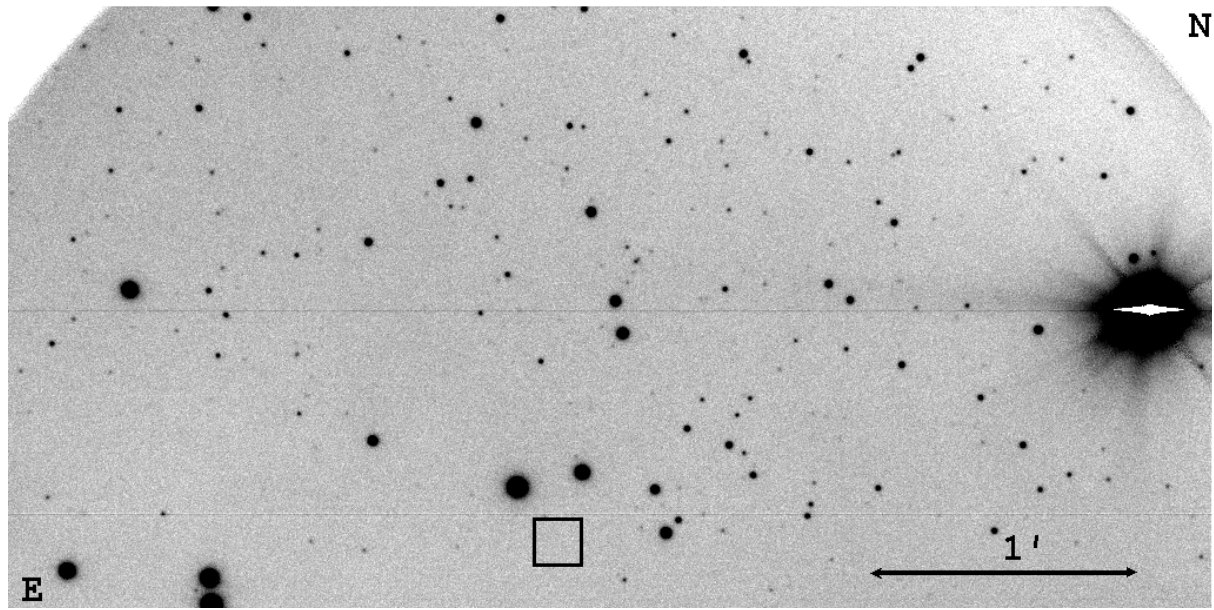}} \\
\end{tabular}
\caption{\label{fov} 
Full-frame FORS2 $v_{\rm HIGH}$-band images of the PSR\, J1809$-$2332 and PSR\,  J1907+0602 fields (upper CCD chip only, 4$\farcm25  \times 2\farcm12$). In each image, the square represents the region reproduced in the two panels of Fig. \ref{fc}.  In both cases, the white regions at the top left and right of the image correspond to the vignetted parts of the detector. In the top panel, the extent of the region covered by the dark nebula Lynds 227 in the upper part of the field of view  is apparent. In the lower panel, the relatively low stellar density in the PSR\,  J1907+0602 field is appreciated.
}
\end{figure*}

\section{Results}

A reassessment of the coordinates of the two pulsars is given in Kerr et al.\ (2015), where they compared those obtained from the improved $\gamma$-ray timing solutions to those obtained from \chan\ observations. For both PSR\, J1809$-$2332 and PSR\,  J1907+0602, the coordinates obtained from $\gamma$-ray timing are affected by a more or less large uncertainty in declination, of  1\farcs32 and 14\farcs41, respectively. Therefore, we assumed for both pulsars the \chan\ coordinates as a reference (see Table \ref{psr}).
We note that, as happens in other cases (e.g. Mignani et al.\ 2016a), the pulsar coordinates reported in {\em Simbad}\footnote{{\texttt http://simbad.u-strasbg.fr/simbad/}} are outdated.
For PSR\, J1809$-$2332, we corrected the reference coordinates for the measured proper motion  (Table \ref{psr}).
to obtain their actual values at the epoch of our VLT observations (MJD=57193). Owing to its faintness in the radio band, no proper motion measurement has been obtained for PSR\,  J1907+0602.

The coordinates of the two pulsars registered on the FORS2 $v_{\rm HIGH}$-band images are shown in Fig.\ref{fc}. No object is detected close to the  PSR\, J1809$-$2332 position, well beyond the uncertainties of our astrometry calibration (Fig.\ref{fc}a). Moreover,  no evidence of extended optical emission that can be associated with the "Taz" PWN is found (Fig.\ref{fov}a).  This is not surprising since the angular scale of the PWN, which stretches $\sim 10\arcmin$ northwest of the pulsar position, is much larger than the field of view of the high-resolution FORS2 images (4$\farcm25  \times 2\farcm12$ for the upper CCD chip), and half of it is covered by the dark nebula Lynds 227 (Oka et al.\ 1999). Therefore, we cannot set significant constraints on the optical surface brightness of the "Taz" PWN.  We note that the coordinates of PSR\, J1809$-$2332 puts the pulsar about 1\arcmin\ south of the edge of  the dark nebula Lynds 227, which 
would denote that the pulsar optical emission is not obscured by the nebular gas.  Surprisingly, in the case of  PSR\,  J1907+0602, no object is detected even within a radius of 5\arcsec\ around the pulsar position (Fig.\ref{fc}b).  This is consistent with the low average density of stellar objects measured in the FORS2 field of view, $\rho \sim 0.07$ arcsec$^{-2}$  (Fig. \ref{fov}b). We note that the PSR\,  J1907+0602 position is a few arcmin away from the Dobashi dark clouds (Dobashi 2011), so that it might be possible that the more teneous clouds' limbs increase the extinction in parts of the FORS2 field of view. This would explain the low average star density.
We note that the possibility that the optical counterpart to PSR\,  J1907+0602 has moved out of the field of view shown in Fig.\ref{fc}b, centred on the pulsar position at the reference epoch (MJD=55555), as the result of its yet unknown proper motion is extremely unlikely. In order to have moved away from the reference position by $\ge 5\arcsec$ in $\sim 4.5$ years, i.e. the time elapsed between the reference epoch and that  of our VLT observations (MJD=57193), the pulsar must have a proper motion $\mu \ge 1\farcs1$ yr$^{-1}$. For its nominal DM distance  $d_{\rm DM}=3.2\pm0.6$ kpc (Abdo et al.\ 2010), this would imply an unrealistic transverse velocity of  $\ge 17600$  ($d_{\rm DM}$/3.2 kpc) km s$^{-1}$.  In order to be compatible with the largest measured pulsar transverse velocities (e.g., Chatterjee \& Cordes 2004), such a large annual displacement would require a pulsar distance $\le 290$ ($d_{\rm DM}$/3.2 kpc) pc. This value is much smaller than one can expect accounting for the uncertainties on the DM-based distance and is also not compatible with the significant $N_{\rm H}$ along the line of sight to the pulsar measured from X-ray observations (see below). Therefore, we can rule out that we could not identify the pulsar counterpart owing to its unknown proper motion. We note that the size of the TeV PWN detected around PSR\,  J1907+0602 is at least $1\degr\times1\degr$, way larger than the field of view of Fig.\ref{fov}b and the same considerations as for the "Taz" PWN also apply in this case.

 \begin{table*}
\begin{center}
\caption{X and $\gamma$-ray spectral parameters and fluxes for the two pulsars discussed in this work (Abdo et al.\ 2013; Acero et al.\ 2015). In the X-rays, only the value of the non-thermal flux ($F_{\rm X} ^{\rm nt}$) is reported.}
\label{spec}
\begin{tabular}{lcccccc} \hline
                                         & $\Gamma_{\rm X}$ & $K T$ &  $F_{\rm X} ^{\rm nt}$ & $\Gamma_{\gamma}$ & $E_c $ & $F_{\gamma}$ \\ 
                                         &                                  & (keV) &  ($10^{-13}$ erg cm$^{-2}$ s$^{-1}$) &                           & (GeV)   & ($10^{-11}$ erg cm$^{-2}$ s$^{-1}$) \\ \hline
PSR\,  J1907+0602         &    $0.93^{+0.15}_{-0.21}$   & -  &  $0.58 \pm 0.14$ &$1.85\pm0.03$ & $4.46\pm0.31$  & $31.90\pm 0.55$  \\
PSR\, J1809$-$2332       &    $1.8\pm0.2$                    &            $ 0.19 \pm0.04$           &  $1.32 \pm 0.30$ & $1.63\pm0.02$ & $3.57\pm0.15$  & $44.79\pm 0.61$ \\  \hline
  \end{tabular}

\end{center}
\end{table*}

With no detected candidate counterpart, both pulsars remain unidentified in the optical domain. We computed the optical flux upper limits at the pulsar positions in both the $v_{\rm HIGH}$  and $b_{\rm HIGH}$  bands from the rms of the sky background (Newberry 1991) measured at the computed pulsar positions.  We computed the flux in an aperture of size equal to the image FWHM and applied the aperture correction computed directly on the images from the growth curve of a number of non-saturated reference stars in the FORS2 field. We applied the airmass  correction as described in Sectn.\, 2. The computed $3\sigma$ magnitude upper limits in the $v_{\rm HIGH}$ and $b_{\rm HIGH}$ bands are $m_{\rm v} \la 27.6$, $m_{\rm b} \la 26.8$ for PSR\, J1809$-$2332 and $m_{\rm v} \la 26.9$, $m_{\rm b} \la 27.7$ for PSR\,  J1907+0602.  The deeper limiting magnitude in the $v_{\rm HIGH}$ band for PSR\, J1809$-$2332 with respect to  PSR\,  J1907+0602, despite the lower exposure time (2700 s against 4500 s; Sectn.\, 3.2) is mainly due to the better image quality ($\sim 0\farcs5$ against $\sim 0\farcs8$).

The value of the hydrogen column density $N_{\rm H}$ towards PSR\, J1809$-$2332 and PSR\,  J1907+0602 is derived from the fits to their X-ray spectra (Abdo et al.\ 2013), and is quite similar: $4.92^{+0.68}_{-0.55} \times 10^{21}$ cm$^{-2}$ and $4.11^{+0.35}_{-0.30} \times 10^{21}$ cm$^{-2}$, respectively.  According to the relation of Predehl\& Schmitt (1995), this corresponds to an interstellar reddening $E(B-V) =0.74^{+0.06}_{-0.05}$ and  $E(B-V) =0.88^{+0.12}_{-0.09}$ for the two pulsars, respectively. We used these values to determine their extinction-corrected optical flux upper limits. For the extinction coefficients  in the $v_{\rm HIGH}$ and $b_{\rm HIGH}$ bands we used the values of Fitzpatrick (1991) as a reference. To be conservative, in both cases we adopted the largest value of the interstellar reddening compatible with the associated uncertainties. The extinction-corrected optical flux limits in the $v_{\rm HIGH}$ and $b_{\rm HIGH}$ bands are $F_{\rm v} \la 2.8\times10^{-16}$ and $F_{\rm b} \la 2.8 \times 10^{-15}$ erg cm$^{-2}$ s$^{-1}$ for PSR\, J1809$-$2332, whereas they are $F_{\rm v} \la 9.4\times10^{-16}$ and $F_{\rm b} \la 2.7\times10^{-15}$ erg cm$^{-2}$ s$^{-1}$  for PSR\,  J1907+0602.

 \begin{table*}
\begin{center}
\caption{Optical flux upper limits in the $v_{\rm HIGH}$ band, optical luminosity, optical efficiency  $\eta_{\rm opt} \equiv L_{\rm opt}/\dot{E}_{\rm rot}$, and optical to X and $\gamma$-ray flux ratios upper limits for the two pulsars discussed in this work.}
\label{opt}
\begin{tabular}{lcccccc} \hline
                                         &  $F_{\rm opt}$ &   $L_{\rm opt}$ &  $ \eta_{\rm opt}$ & $F_{\rm opt}/F_{\rm X}^{\rm nt}$ & $F_{\rm opt}/F_{\gamma}$  \\ 
                                         &   ($10^{-16}$ erg cm$^{-2}$ s$^{-1}$) & ($10^{30}$ erg s$^{-1}$) &  & & & \\ \hline
PSR\,  J1907+0602         &    $\la 9.4$   & $\la 1.63$ &  $\la 5.8 \times 10^{-7}$ &$\la1.3\times10^{-2}$ & $\la 2.9 \times 10^{-6}$ \\
PSR\, J1809$-$2332       &    $\la 2.8$   & $\la 0.24$ &  $ \la 5.7 \times 10^{-7}$ &$\la1.7\times10^{-3}$ &$\la 6.2 \times 10^{-7}$ \\ \hline
  \end{tabular}

\end{center}
\end{table*}
\section{Discussion}

We compared the unabsorbed optical flux upper limits at the peaks of the $v_{\rm HIGH}$ and $b_{\rm HIGH}$ bands with the extrapolation in the optical regime of the models best-fitting the X and $\gamma$-ray spectra of the two pulsars. As explained in Sectn.\, 1, since they are both relatively young (characteristic age 19.5 and 67.6 kyr), we expect that any optical emission is mainly ascribed to non-thermal processes in the neutron star magnetosphere, as observed in the Vela pulsar. Therefore, we focus our comparison on the non-thermal emission components.  The X-ray spectrum of PSR\,  J1907+0602  is fit by a single PL model with photon index $\Gamma_{\rm X} = 0.93^{+0.15}_{-0.21}$, whereas for  the slightly older PSR\, J1809$-$2332 (67.6 kyr) the best-fit requires the addition of a blackbody (BB) component with $kT = 0.19 \pm0.04$ keV to the PL spectrum ($\Gamma_{\rm X} = 1.8\pm0.2$), produced from hot spots on the neutron star surface (black body radius $R_{BB}=1.5^{+1.26}_{-0.44}$ km for a distance of 1.7 kpc), with an unabsorbed flux $F_{\rm X}^{\rm BB} = (1.75 \pm 1.3) \times 10^{-13}$  erg cm$^{-2}$ s$^{-1}$. The unabsorbed non-thermal X-ray flux $F_{\rm X}^{\rm nt}$  in the 0.3--10 keV band is $(0.58 \pm 0.14) \times 10^{-13}$  and  $(1.32 \pm 0.30) \times 10^{-13}$ erg cm$^{-2}$ s$^{-1}$ for PSR\,  J1907+0602 and PSR\, J1809$-$2332, respectively (Abdo et al.\ 2013).  The $\gamma$-ray spectra of the two pulsars are described by a PL with an exponential cut off (Acero et al.\ 2015), where the PL photon index $\Gamma_{\gamma}=1.85\pm0.03$ and the cut-off energy $E_{\rm c}= 4.46\pm0.31$ GeV for PSR\,  J1907+0602, whereas $\Gamma_{\gamma}=1.63\pm0.02$ and $E_{\rm c}= 3.57\pm0.15$ GeV for PSR\, J1809$-$2332. These parameters correspond to 0.1--100 GeV  fluxes $F_{\gamma}$ of $(31.9\pm 0.55) \times 10^{-11}$ and  $(44.79\pm 0.61) \times 10^{-11}$ erg cm$^{-2}$ s$^{-1}$, respectively. The X/$\gamma$-ray spectral parameters and the values of the corresponding non-thermal fluxes of the two pulsars are summarised in Table \ref{spec}.

\begin{figure}
\centering
\begin{tabular}{cc}
\subfloat[PSR\, J1907+0602]{\includegraphics[width=8.9cm, clip=]{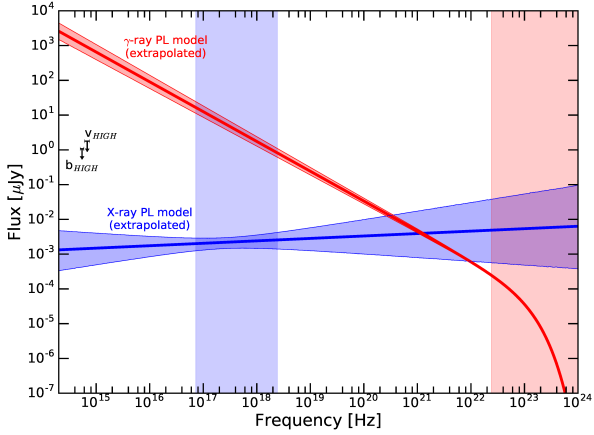}} \\
\subfloat[PSR\, J1809-2332]{\includegraphics[width=8.9cm, clip=]{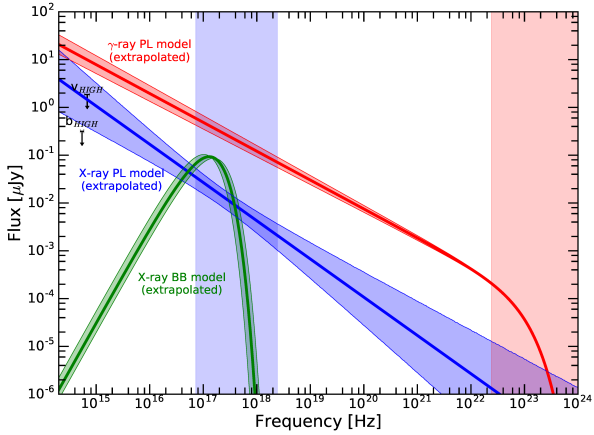}} 
\end{tabular}
\caption{\label{sed} Spectral energy distribution for PSR\,  J1907+0602   and  PSR\, J1809$-$2332. The optical flux upper limits are labelled with the filter names. The blue and red lines represent the extrapolation in the  optical regime of the PLs best-fitting the X and $\gamma$-ray spectra, respectively, whereas the green curve in the lower panel corresponds to the BB component to the X-ray spectrum of PSR\, J1809$-$2332. The  filled regions defined by the  dashed lines correspond to the $1 \sigma$ error. The vertical blue and red rectangles indicate the spectral energy range covered by the X-ray and $\gamma$-ray observations, i.e. 0.3--10 keV and $\ge$100 MeV, respectively.
}
\end{figure}

In the optical, the unabsorbed flux upper limits are 0.31 and 1.86 $\mu$Jy for PSR\, J1809$-$2332, whereas they are 1.07 and 1.77 $\mu$Jy for PSR\,  J1907+0602, computed at frequencies $\nu_v= 5.45 \times 10^{5}$ GHz and  $\nu_b = 6.81 \times 10^{5}$ GHz, respectively, corresponding to the peak frequencies of the $v_{\rm HIGH}$ and $b_{\rm HIGH}$ bands. The results are shown in Fig.\ref{sed}. In the case of PSR\,  J1907+0602, the optical flux upper limits lie between the extrapolations of the X and $\gamma$-ray PLs, which confirms a turn-over in the $\gamma$-ray spectrum at low energies,  as already implied by the extrapolation of the X-ray PL. The optical emission could be compatible with the extrapolation of the X-ray PL only if the actual pulsar optical fluxes were three orders of magnitude fainter than our upper limits. For PSR\, J1809$-$2332  the optical flux upper limits are below the extrapolation of the $\gamma$-ray PL.  Therefore, also in this case the $\gamma$-ray spectrum features a turn-over at low energies, which is also observed in other pulsars (see Mignani et al.\ 2016a and references therein).  The  $b_{\rm HIGH}$-band flux upper limit is just below the $1\sigma$ uncertainty on the  extrapolation of the X-ray PL, and might indicate the existence of another spectral break, between the optical and the X-rays.
 In the thermal regime, the extrapolation of the X-ray BB predicts very faint optical fluxes, as expected from the small emitting area and high temperature of the hot spots on the neutron star surface. Any comparison with a BB spectrum produced from the rest of the neutron star surface, presumably at a lower temperature than the hot spots, cannot be discussed yet owing to the lack of temperature constraints from X-ray observations.  However, thermal radiation from the bulk of the neutron star surface usually contributes significantly to the optical emission for pulsars older than 100 kyrs (Mignani 2011), so that its contribution is expected to be small in the case of PSR\, J1809$-$2332.
In general, the picture that emerges from the spectral energy distributions (SEDs) of the two pulsars is that the optical emission is not obviously related to the X or $\gamma$-ray one.  Furthermore, multi-wavelength non-thermal emission spectra of pulsars can be quite different from case to case, even for objects of comparable age, as also discussed, based on a broader sample, by Mignani et al.\ (2016a). This means that it is difficult to track a clear SED evolution, at least in the age range from a few to a few tens of kyr. What causes the difference in the SEDs is still unclear although it is likely that this is related to a difference in the particle density and energy distribution in the magnetosphere regions where the emission at different wavelengths is generated, as well as to a more or less favourable viewing angle to these regions.

We used the unabsorbed optical flux upper limits in the $v_{\rm HIGH}$ band to determine the upper limits on the optical luminosity $L_{\rm opt}$ and on the optical emission efficiency $\eta_{\rm opt}$ defined as the fraction of the rotational energy loss rate converted into optical luminosity, $\eta_{\rm opt} \equiv L_{\rm opt}/\dot{E}_{\rm rot}$, for the two pulsars. Similarly, we used the unabsorbed optical flux upper limits to constrain the pulsar multi-wavelength properties from the ratios between their unabsorbed, non-thermal fluxes, $F_{\rm opt}/F_{\rm X}^{\rm nt}$ and $F_{\rm opt}/F_{\gamma}$. In all cases, we conservatively accounted for the uncertainty on the pulsar distance to compute the upper limits on $L_{\rm opt}$ and $\eta_{\rm opt}$ and for the uncertainties on the X and $\gamma$-ray fluxes $F_{\rm X}^{\rm nt}$ and $F_{\gamma}$ to compute the upper limits on the $F_{\rm opt}/F_{\rm X}^{\rm nt}$ and $F_{\rm opt}/F_{\gamma}$ ratios.
For the Vela-like pulsar PSR\,  J1907+0602 ($\tau_c = 19.5$ kyr), we derived $L_{\rm opt} \la 1.63 \times 10^{30}$ erg s$^{-1}$ and $\eta_{\rm opt} \la 5.8 \times 10^{-7}$, $F_{\rm opt}/F_{\rm X}^{\rm nt}\la0.013$, and $F_{\rm opt}/F_{\gamma} \la 2.9 \times 10^{-6}$.  We compared these values with the corresponding ones for the 11.2 kyr-old Vela $\gamma$-ray pulsar (e.g., Moran et al.\ 2013).  As far as the optical luminosity is concerned, Vela has $L_{\rm opt} \sim 10^{28}$ erg s$^{-1}$, which makes it the least luminous of the young ($\tau_{\rm c} \la 100$ kyr) $\gamma$-ray pulsars detected in the optical.  Vela has a rotational energy loss rate about twice as large as PSR\,  J1907+0602 and an optical emission efficiency $\eta_{\rm opt} \sim1.9 \times 10^{-9}$, the lowest of all $\gamma$-ray pulsars detected in the optical (Moran et al.\ 2013), regardless of their characteristic age. The upper limits on $L_{\rm opt} $ and $\eta_{\rm opt}$ obtained for  PSR\,  J1907+0602 and a handful of other $\gamma$-ray pulsars of comparable characteristic age and with comparably deep detection limits  (Mignani et al.\ 2016a) are all above the values measured for the Vela pulsar, owing to the much larger distance to these pulsars and the much larger  interstellar extinction  along the line of sight.
 For these pulsars, the limits on $L_{\rm opt}$ and $\eta_{\rm opt}$ are also well below those obtained for very young $\gamma$-ray pulsars ($\tau_c \la 5$ kyr), such as the Crab pulsar, which has an optical luminosity $L_{\rm opt} \approx 10^{33}$ erg s$^{-1}$ and an emission efficiency $\eta_{\rm opt} \approx  10^{-5}$. This shows that around a characteristic age of 10 kyr the pulsar optical emission efficiency 
tends to  decrease. Whether all 10 kyr-ish old pulsars are as low efficient optical emitters as Vela is, however, unclear yet owing to the lack of detections of pulsars in this crucial age range. Owing to the relatively low unabsorbed X-ray flux of PSR\,  J1907+0602 
the limit on the $F_{\rm opt}/F_{\rm X}^{\rm nt}$ is less constraining than those obtained for other pulsars of comparable age (Mignani et al.\ 2016a), whereas the limit on the $F_{\rm opt}/F_{\gamma}$ flux ratio is comparable to those obtained for most of them, accounting for the large scatter in the value distribution and for the uncertainties in the actual beaming factors in the optical and $\gamma$ rays. 
Similarly as above, for the slightly older PSR\, J1809$-$2332 ($\tau_c = 67.6$ kyr) we derived $L_{\rm opt} \la 0.24 \times 10^{30}$ erg s$^{-1}$, $\eta_{\rm opt} \la 5.7 \times 10^{-7}$, $F_{\rm opt}/F_{\rm X}^{\rm nt}\la1.7\times10^{-3}$, and $F_{\rm opt}/F_{\gamma} \la 6.2 \times 10^{-7}$.  The limit on $\eta_{\rm opt}$ is comparable to that obtained for PSR\,  J1907+0602, in line with the trend of a decrease of $\eta_{\rm opt}$ with the pulsar characteristic age.  The limits on the 
$F_{\rm opt}/F_{\rm X}^{\rm nt}$ and $F_{\rm opt}/F_{\gamma}$  ratios are more constraining than for PSR\,  J1907+0602, owing to the larger X and $\gamma$-ray fluxes and are also in line with those reported in Mignani et al.\ (2016a).  The limits obtained for both pulsars (summarised in Table \ref{opt}) confirm that, in general, the $F_{\rm opt}/F_{\rm X}^{\rm nt}$  tends to be never above $10^{-3}$ and the $F_{\rm opt}/F_{\gamma}$ never above $10^{-6}$ for pulsars in the 10--100 kyr age range. This is a useful piece of information to estimate upper limits on the expected optical fluxes when planning optical follow-up of X and $\gamma$-ray pulsars.

As shown above, our observations support the scenario where pulsars in the age range 10--100 kyrs have a low efficiency  in the conversion of rotational energy into optical radiation with respect to younger pulsars (see also Mignani et al.\ 2016a)  and, as such, they are more elusive targets in the optical band, in spite of their relatively high energy budget ($\dot{E}_{\rm rot} \approx 10^{35}$--$10^{37}$ erg s$^{-1}$). Whether this traces an evolution in the optical emission efficiency $\eta_{\rm opt}$ with the characteristic age is unclear. At least in some cases, the lack of a detection might be explained by an unfavourable beaming/viewing geometry in the optical or by a very steep optical spectrum.
Interestingly, the very young ($\tau_c \sim 5$ kyr) pulsar PSR\, J0537$-$6910 in the Large Magellanic Cloud, which has the highest $\dot{E}_{\rm rot}$ in the pulsar family ($4.9 \times10^{38}$ erg s$^{-1}$),  is yet undetected in the optical/near ultraviolet in spite of deep observations (Mignani et al.\ 2000; 2005; 2007). This implies an $\eta_{\rm opt} \la 1.2 \times 10^{-7}$, much lower than the 1.2 kyr-old Crab pulsar ($\eta_{\rm opt}  \approx 10^{-5}$) and comparable to the limits obtained for pulsars of age comparable to Vela (e.g., Mignani et al.\ 2016a). This might suggest that the $\eta_{\rm opt}$ evolution, if indeed occurs, is more sudden than expected. Optical investigations of pulsars younger than 10 kyr would be crucial to clarify this issue. Unfortunately, for many of them the large distance and interstellar extinction severely affect the chances of an optical detection.

\section{Summary and conclusions}

We observed the two $\gamma$-ray pulsars PSR\,  J1907+0602 and  PSR\, J1809$-$2332 with the VLT.  With them, there are now six  isolated $\gamma$-ray pulsars discovered by {\em Fermi}, for which the VLT obtained the first deep optical observations. The others are: PSR\, J1357$-$6429 (Mignani et al.\ 2011), PSR\, J1028$-$5819 (Mignani et al.\ 2012), PSR\, J1048$-$5832 (Razzano et al.\  2013; Danilenko et al.\ 2013), and PSR\, J1741$-$2054 (Mignani et al.\ 2016b). Of them, a candidate optical counterpart has been found for PSR\, J1741$-$2054 (Mignani et al.\ 2016b), whereas for PSR\, J1357$-$6429  a candidate counterpart might have been identified in the near infrared (Zyuzin et al.\ 2016). Thus, one can certainly say that the VLT leads the optical follow-up of isolated $\gamma$-ray pulsars, at least in the southern hemisphere. The VLT yielded the only optical counterpart identifications obtained so far for the new southern $\gamma$-ray pulsars discovered by {\em Fermi}\footnote{The optical counterpart to PSR\, B0540$-$69  in the Large Magellanic Cloud (Caraveo et al.\ 1992) was identified long before its detection as a $\gamma$-ray pulsar (Ackermann et al.\ 2015).}. In the northern hemisphere, a candidate optical counterpart to the $\gamma$-ray pulsar PSR\, J0205+6449 was discovered in archival data taken with the Gemini telescope (Moran et al.\ 2013).
Unfortunately, in this case we were not successful in detecting the optical counterparts to the target pulsars,  and we could only set $3\sigma$ upper limits to their optical brightness 
of $m_{\rm v} \sim 26.9$ and $m_{\rm v} \sim 27.6$ for PSR\,  J1907+0602 and  PSR\, J1809$-$2332,
respectively. These are the deepest constraints on the optical fluxes ever obtained for these two pulsars, which had never been observed with 10m-class telescopes so far.

\section*{Acknowledgments}
We thank the referee, Stephen Chi-Yung Ng, for his constructive comments to our manuscript. RPM acknowledges financial support from the project TECHE.it. CRA 1.05.06.04.01 cap 1.05.08 for the project "Studio multilunghezze d'onda da stelle di neutroni con particolare riguardo alla emissione di altissima energia". The work of MM was supported by the ASI-INAF contract I/037/12/0, art.22 L.240/2010 for the project $"$Calibrazione ed Analisi del satellite NuSTAR$"$. AB and DS acknowledge support through
EXTraS, funded from the European Union's Seventh Framework Programme for research,
technological development and demonstration under grant agreement no 607452.

\label{lastpage}

\end{document}